\documentstyle[aps,epsf,eqsecnum,feynmp
]{revtex}
\newcommand{\setval}{\fmfset{wiggly_len}{1.5mm}
\fmfset{arrow_len}{1.5mm}\fmfset{arrow_ang}{13}\fmfset{dash_len}{1.5mm}
\fmfpen{0.125mm}\fmfset{dot_size}{1thick}}
\fmfset{dot_size}{1thick}
\begin{document}
\begin{fmffile}{weissbach}
\title{Variational Perturbation Theory for
the Ground-State Wave Function}

\author{A.~Pelster and F.~Weissbach}

\address{Institut f\"ur Theoretische Physik, Freie Universit\"at Berlin, \\
Arnimallee 14, D-14195 Berlin, Germany\\
E-mails: pelster@physik.fu-berlin.de, florian.weissbach@physik.fu-berlin.de}
\maketitle
\begin{abstract}
We evaluate perturbatively 
the density matrix\index{density matrix}
in the low-temperature limit
and thus the ground-state wave function\index{wave function}
of the anharmonic oscillator\index{anharmonic oscillator}
up to second order in the coupling constant. We then employ Kleinert's 
variational perturbation theory
\index{variational perturbation theory}
to determine the ground-state
wave function for all coupling strengths.
\end{abstract}
\section{Introduction}
Variational perturbation theory\index{variational perturbation theory}
as developed by Kleinert\index{KLEINERT, H.} \cite{Kleinert}
provides a systematic algorithm to evaluate perturbation series
at all coupling strengths
including the strong-coupling limit $g \rightarrow \infty$.
It was thoroughly investigated
for the ground-state energy of the anharmonic oscillator
up to the 150th order \cite{Janke1,Janke2}, 
and its convergence was found to be exponentially fast and uniform.
A similar systematic study has not yet been performed for the
ground-state wave function. A first-order variational approach
was set up by 
Kunihiro\index{KUNIHIRO, T.} \cite{Kunihiro}.
However, his method did not satisfactorily deal with certain problems
in the variational procedure which will be discussed in detail below.

In this work we improve Kunihiro's\index{KUNIHIRO, T.} 
first-order calculation  and
extend the treatment to the second order in the coupling strength.
The ground-state wave function of the anharmonic oscillator
is calculated
from the low-temperature limit of the diagonal elements of the
density matrix\index{density matrix}.
A variational evaluation of
the density matrix for the double-well
potential has already been perfomed for finite temperatures
in Ref. \cite{Dichtematrix}.
\section{Perturbation Theory\index{perturbation theory}}
Consider a quantum-mechanical point particle of mass $M$ moving
in the one-dimensional anharmonic oscillator potential 
\begin{eqnarray}
\label{AHOpotential}
V(x) = \frac{M}{2}  \omega^2 x^2 + g x^4 \, ,
\end{eqnarray}
where $\omega$ denotes the frequency 
and $g$ the coupling constant.
We determine its ground-state wave function $\Psi (x)$ by
evaluating the low-temperature limit of the
diagonal elements of the density matrix\index{density matrix}
\begin{eqnarray}
\label{wavefunction}
\Psi (x) = \lim_{\beta \rightarrow \infty} \sqrt{\rho (x,x)}\, ,
\end{eqnarray}
which is defined by
\begin{eqnarray}
\label{densitymatrix}
\rho (x_b,x_a) = \frac{(x_b \, \hbar \beta | x_a \, 0)}{Z}\, .
\end{eqnarray}
Here $(x_b \, \hbar \beta | x_a \, 0)$ denotes the imaginary-time 
evolution amplitude \index{time evolution amplitude}
with the path integral representation \cite{Kleinert}
\begin{eqnarray}
(x_b \, \hbar \beta | x_a \, 0) = 
\int_{x(0)=x_a}^{x(\hbar \beta)=x_b} {\cal D} x   
\, \exp \left\{- \frac{1}{\hbar} \int_0^{\hbar \beta} d \tau
\left[ \frac{M}{2} \dot{x}^2(\tau) 
+\frac{M}{2} \omega^2 x^2(\tau) + g x^4(\tau) \right] \right\} \, ,
\label{timeevolutionpathintegral} 
\end{eqnarray}
and $Z$ denotes the partition function\index{partition function}
\begin{eqnarray}
\label{partitionfunction}
Z = \int_{- \infty}^{+ \infty} dx \, (x \, \hbar \beta | x \,  0) \, .
\end{eqnarray}
By expanding Eq.~(\ref{timeevolutionpathintegral}) in powers of the
coupling constant $g$ we obtain the
perturbation series\index{perturbation series}
\begin{eqnarray}
\label{2B}
(x_b \, \hbar \beta | x_a \, 0)   =  
(x_b \, \hbar \beta | x_a \, 0)_{\omega}
\left[ 1 - \frac{g}{\hbar} \int_0^{\hbar\beta} d \tau_1 
\langle x^4 (\tau_1) \rangle_{\omega}
\, + \, 
\frac{g^2}{2 \hbar^2} \int_0^{\hbar \beta} d \tau_1
\int_0^{\hbar \beta} d \tau_2 \, 
\langle x^4 (\tau_1) \, x^4 (\tau_2) \rangle_{\omega}
+ \, ... \right] \, ,
\end{eqnarray}
where we have introduced the harmonic imaginary-time evolution
amplitude\index{time evolution amplitude}
\begin{eqnarray}
(x_b \, \hbar \beta | x_a \, 0)_{\omega} \equiv 
\int_{x(0)=x_a}^{x(\hbar \beta)=x_b} 
{\cal D} x  
\, \exp \left\{ -\frac{1}{\hbar} \int_{0}^{\hbar \beta} d \tau
\left[ \frac{M}{2} \dot{x}^2 (\tau) + \frac{M}{2} \omega^2 x^2 (\tau) 
\right] \right\} \, , 
\label{harmonic}
\end{eqnarray}
and the harmonic expectation value\index{expectation value}
for an arbitrary functional $F[x]$ of the path
$x(\tau)$:
\begin{eqnarray}
\label{expectation}
\langle F[x] \rangle_{\omega}  \equiv 
\frac{1}{(x_b \, \hbar \beta | x_a \, 0)_{\omega} }
\int_{x(0)=x_a}^{x(\hbar \beta)=x_b} {\cal D} x 
\, F[x] 
\, \exp \left\{ - \frac{1}{\hbar} \int_0^{\hbar \beta} d \tau
\left[ \frac{M}{2} \dot{x}^2(\tau)
+ \frac{M}{2} \omega^2 x^2(\tau) \right] \right\} \, .
\end{eqnarray}
The latter is evaluated with the help of
the generating functional\index{generating functional}
for the harmonic oscillator, whose path integral representation reads
\begin{eqnarray}
\label{timeevolution}
(x_b \, \hbar \beta | x_a \, 0)_{\omega} [j]  
= \int_{x(0)=x_a}^{x(\hbar \beta)=x_b}
{\cal D} x  
\exp \left\{ - \frac{1}{\hbar} \int_{0}^{\hbar \beta} d \tau  
\, \left[ \frac{M}{2} \dot{x}^2 (\tau)
+ \frac{M}{2} \omega^2 x^2 (\tau) - j(\tau) x(\tau) \right]\right\} \, ,
\end{eqnarray}
leading to \cite{Kleinert}
\begin{eqnarray}
(x_b \, \hbar \beta | x_a \, 0)_{\omega} [j]  = 
(x_b \, \hbar \beta | x_a \, 0)_{\omega} \exp \left[
\frac{1}{\hbar} \int_0^{\hbar \beta} d \tau_1 
\, x_{\rm cl} (\tau_1) j(\tau_1) 
\, + \frac{1}{2 \hbar^2} \int_0^{\hbar \beta} d \tau_1 \int_0^{\hbar \beta} 
d \tau_2 \,\,  G (\tau_1 , \tau_2) j(\tau_1) j(\tau_2) \right] 
\label{timeevolution2}
\end{eqnarray}
with
\begin{eqnarray}
(x_b  \, \hbar \beta |  x_a \,  0)_{\omega}
= \sqrt { \frac{M \omega}{2 \pi \hbar \sinh \hbar \beta \omega} }  
\, \exp \left\{ - \frac{M \omega}{2 \hbar \sinh \hbar \beta \omega}
[(x_a^2+x_b^2) 
\cosh \hbar \beta \omega - 2 x_a x_b] \right\} \, .
\label{harmonic2}
\end{eqnarray}
In Eq. (\ref{timeevolution2})
we have introduced the classical path\index{classical path}
\begin{eqnarray}
\label{xcl}
x_{\rm cl} (\tau ) \equiv \frac{x_a \sinh(\hbar \beta - \tau ) \omega 
+ x_b \sinh \omega \tau}{\sinh \hbar \beta \omega} \, ,
\end{eqnarray}
and the Green function\index{Green function}
\begin{eqnarray}
\label{Green}
G (\tau_1 , \tau_2) \equiv   
\frac{\hbar}{2 M \omega}
\frac{1}{\sinh \hbar \beta \omega}  
\left[ \theta(\tau_1 - \tau_2) \sinh (\hbar \beta - \tau_1) \omega 
\sinh \omega \tau_2 
\, + \, \theta(\tau_2 - \tau_1) \sinh (\hbar \beta - \tau_2)\omega
\sinh \omega \tau_1 \right] \, .
\end{eqnarray}
We follow Ref. \cite{KPB} and evaluate harmonic expectation values
of polynomials in $x$ 
arising from the generating functional (\ref{timeevolution2})
according to Wick's theorem\index{Wick theorem}. 
Let us illustrate
the procedure to reduce the power of the polynomial
by the example of the harmonic expectation value
\begin{eqnarray}
\label{wick}
\langle x^n(\tau_1) \, x^m(\tau_2) \rangle_{\omega} \, .
\end{eqnarray}
\renewcommand{\labelenumi}{(\roman{enumi})}
\begin{enumerate}
\item Contracting $x (\tau_1)$ with $x^{n-1}(\tau_1)$
and $x^m (\tau_2)$ leads to a
Green function $G (\tau_1, \tau_1)$ and $G (\tau_1, \tau_2)$
with multiplicity
$n-1$ and $m$, respectively. The rest of the polynomial
remains within the harmonic expectation value, leading to
$\langle x^{n-2}(\tau_1) \, x^m (\tau_2) \rangle_{\omega}$ and
$\langle x^{n-1}(\tau_1) \, x^{m-1} (\tau_2) \rangle_{\omega}$.
\item If $n>1$, extract one $x(\tau_1)$ from the expectation value
giving $x_{\rm cl}(\tau_1)$ multiplied
by $ \langle x^{n-1}(\tau_1) x^m (\tau_2) \rangle_{\omega}$.
\item Add the terms (i) and (ii).
\item Repeat the previous steps 
until only products of expectation values
$ \langle x (\tau_1)\rangle_{\omega}
= x_{\rm cl} (\tau_1)$ remain. 
\end{enumerate}
With the help of this procedure,
the first-order harmonic expectation value 
$\langle x^4(\tau_1) \rangle_{\omega}$ is reduced to
\begin{eqnarray}
\label{newwickstheorem}
\langle x^4 (\tau_1) \rangle_{\omega} 
= x_{\rm cl} (\tau_1) \, \langle x^3 (\tau_1)  \rangle_{\omega}
+ 3 \, G (\tau_1 , \tau_1) \langle \, x^2 (\tau_1) \rangle_{\omega} \, .
\end{eqnarray}
Furthermore, we find
\begin{eqnarray}
\label{egal}
\langle x^3 (\tau_1) \rangle_{\omega} 
= x_{\rm cl} (\tau_1) \langle x^2 (\tau_1) 
\rangle_{\omega} 
+ 2 G (\tau_1, \tau_1) \, x_{\rm cl} (\tau_1) \, ,
\end{eqnarray}
and
\begin{eqnarray}
\label{egal2}
\langle x^2 (\tau_1) \rangle_{\omega} = x^2_{\rm cl} (\tau_1) 
+ G (\tau_1, \tau_1) \, .
\end{eqnarray}
Combining Eqs. (\ref{newwickstheorem})--(\ref{egal2}) we obtain
in first order
\begin{eqnarray}
\label{x4cl}
\langle x^4 (\tau_1) \rangle_{\omega} 
& = & x^4 _{\rm cl} (\tau_1) \,
+ 6 \, x^2 _{\rm cl} (\tau_1) \, G (\tau_1 , \tau_1) 
+ 3 \, G^2 (\tau_1, \tau_1) \, .
\end{eqnarray}
The second-order harmonic expectation value requires
considerably more effort and finally leads to
\begin{eqnarray}
\langle x^4 (\tau_1) \, x^4 (\tau_2) \rangle_{\omega} 
& = & x^4_{\rm cl} (\tau_1) \, x^4_{\rm cl} (\tau_2) 
+ 16 \, x^3_{\rm cl} (\tau_1) \, G (\tau_1, \tau_2) \, x^3_{\rm cl} 
(\tau_2) 
+ 12 \, x^2_{\rm cl} (\tau_1) \, G (\tau_1, \tau_1) \, 
x^4_{\rm cl} (\tau_2)
\nonumber \\ & & 
+ 72 \, x^2_{\rm cl} (\tau_1) \, G^2 (\tau_1, \tau_2) \, x^2_{\rm cl} 
(\tau_2) 
+ 36 \, x^2_{\rm cl} (\tau_1) \, G (\tau_1, \tau_1) \, G (\tau_2, \tau_2) 
\, x^2_{\rm cl} (\tau_2) 
+ 9 \, G^2 (\tau_1, \tau_1) \, G^2 (\tau_2, \tau_2) 
\nonumber \\ & & 
+ 96 \, x^3_{\rm cl} (\tau_1) \, 
G (\tau_1, \tau_2) \, G (\tau_2, \tau_2) 
\, x_{\rm cl} (\tau_2) 
+ 6 \, G^2 (\tau_1, \tau_1) \, x^4_{\rm cl} (\tau_2) 
+ 96 \, x_{\rm cl} (\tau_1) \, G^3 (\tau_1, \tau_2) \, x_{\rm cl} (\tau_2) 
\nonumber \\ & & 
+ 144 \, x_{\rm cl} (\tau_1) \, G (\tau_1, \tau_1) \, 
G (\tau_1, \tau_2) \, 
G (\tau_2, \tau_2) \, x_{\rm cl} (\tau_2) 
+ 36 \, G^2 (\tau_1, \tau_1) \, x^2_{\rm cl} (\tau_2) \, 
G (\tau_2, \tau_2) 
\nonumber \\ & & 
+ 144 \, x^2_{\rm cl} (\tau_1) \, G^2 (\tau_1, \tau_2) \, 
G (\tau_2, \tau_2) 
+ 72 \, G (\tau_1, \tau_1) \, G^2 (\tau_1, \tau_2) \, G (\tau_2, \tau_2)  
+ 24 \, G^4 (\tau_1, \tau_2) \, .
\label{expectation2}
\end{eqnarray}
The contractions can be illustrated
by Feynman diagrams\index{Feynman diagrams} with the
following rules.
A vertex represents the integration over $\tau$
\begin{eqnarray}
\label{vertex}
\parbox{5mm}{\centerline{
\begin{fmfgraph*}(5,5)
\setval
\fmfforce{0w,1h}{v1}
\fmfforce{1w,1h}{v2}
\fmfforce{0w,0h}{v3}
\fmfforce{1w,0h}{v4}
\fmfforce{1/2w,1/2h}{v5}
\fmf{plain}{v1,v4}
\fmf{plain}{v2,v3}
\fmfdot{v5}
\end{fmfgraph*}}}
\hspace*{3mm} = \hspace*{2mm} \int_0^{\hbar \beta} d \tau \, ,
\end{eqnarray}
a line denotes the Green function\index{Green function}
\begin{eqnarray}
\label{line}
\parbox{10mm}{\centerline{
\begin{fmfgraph*}(10,10)
\setval
\fmfforce{0w,1/2h}{v1}
\fmfforce{1w,1/2h}{v2}
\fmf{plain}{v1,v2}
\fmfv{decor.size=0, label=${\scriptstyle 1}$, l.dist=1mm, l.angle=-180}{v1}
\fmfv{decor.size=0, label=${\scriptstyle 2}$, l.dist=1mm, l.angle=0}{v2}
\end{fmfgraph*}}}
\hspace*{5mm} = \hspace*{2mm} G (\tau_1, \tau_2) \, ,
\end{eqnarray}
and a cross pictures a classical path
\begin{eqnarray}
\label{cross}
\parbox{10mm}{\centerline{
\begin{fmfgraph*}(10,10)
\setval
\fmfforce{0w,1/2h}{v1}
\fmfforce{1w,1/2h}{v2}
\fmf{plain}{v1,v2}
\fmfv{decor.shape=cross,decor.filled=shaded,decor.size=3thick}{v1}
\fmfv{decor.size=0, label=${\scriptstyle 1}$, l.dist=1mm, l.angle=0}{v2}
\end{fmfgraph*}}}
\hspace*{5mm} = \hspace*{2mm} x_{\rm cl} (\tau_1) \, .
\end{eqnarray}
Inserting the harmonic expectation values (\ref{x4cl}) and 
(\ref{expectation2}) into the perturbation expansion (\ref{2B}) leads
in first order to the diagrams 
\begin{eqnarray}
\label{firstorderfeynman}
%
\int_0^{\hbar \beta} d \tau_1 \langle x^4 (\tau_1) \rangle_{\omega} 
\hspace*{0.3cm} \equiv \hspace*{0.3cm}
%
\parbox{10mm}{\centerline{
\begin{fmfgraph*}(10,14)
\setval
\fmfforce{1/2w,12/14h}{v1}
\fmfforce{0w,1/2h}{v2}
\fmfforce{1/2w,1/2h}{v3}
\fmfforce{1w,1/2h}{v4}
\fmfforce{1/2w,2/14h}{v5}
\fmf{plain}{v1,v5}
\fmf{plain}{v2,v4}
\fmfdot{v3}
\fmfv{decor.shape=cross,decor.filled=shaded,decor.size=3thick}{v1}
\fmfv{decor.shape=cross,decor.filled=shaded,decor.size=3thick}{v2}
\fmfv{decor.shape=cross,decor.filled=shaded,decor.size=3thick}{v4}
\fmfv{decor.shape=cross,decor.filled=shaded,decor.size=3thick}{v5}
\end{fmfgraph*}}}
\hspace*{3mm} + 6 \hspace*{3mm}
\parbox{10mm}{\centerline{
\begin{fmfgraph*}(10,10)
\setval
\fmfforce{1/2w,1h}{v1}
\fmfforce{0w,1/2h}{v2}
\fmfforce{1/2w,1/2h}{v3}
\fmfforce{1w,1/2h}{v4}
\fmf{plain}{v2,v4}
\fmf{plain,left=1}{v3,v1,v3}
\fmfdot{v3}
\fmfv{decor.shape=cross,decor.filled=shaded,decor.size=3thick}{v2}
\fmfv{decor.shape=cross,decor.filled=shaded,decor.size=3thick}{v4}
\end{fmfgraph*}}}
\hspace*{3mm} + 3 \hspace*{3mm}
\parbox{10mm}{\centerline{
\begin{fmfgraph*}(10,10)
\setval
\fmfforce{0w,1/2h}{v1}
\fmfforce{1/2w,1/2h}{v2}
\fmfforce{1w,1/2h}{v3}
\fmf{plain,left=1}{v1,v2,v1}
\fmf{plain,left=1}{v2,v3,v2}
\fmfdot{v2}
\end{fmfgraph*}}}
\hspace*{3mm} ,
\end{eqnarray}
whereas the second-order terms are
\begin{eqnarray}
\label{secondorderfeynman}
&&
%
\int_0^{\hbar \beta} d \tau_1 \int_0^{\hbar \beta} d \tau_2 \, 
\langle x^4 (\tau_1) \, x^4 (\tau_2) \rangle_{\omega}  
\hspace*{0.3cm} \equiv \hspace*{0.3cm}
%
\parbox{10mm}{\centerline{
\begin{fmfgraph*}(10,14)
\setval
\fmfforce{1/2w,12/14h}{v1}
\fmfforce{0w,1/2h}{v2}
\fmfforce{1/2w,1/2h}{v3}
\fmfforce{1w,1/2h}{v4}
\fmfforce{1/2w,2/14h}{v5}
\fmf{plain}{v1,v5}
\fmf{plain}{v2,v4}
\fmfdot{v3}
\fmfv{decor.shape=cross,decor.filled=shaded,decor.size=3thick}{v1}
\fmfv{decor.shape=cross,decor.filled=shaded,decor.size=3thick}{v2}
\fmfv{decor.shape=cross,decor.filled=shaded,decor.size=3thick}{v4}
\fmfv{decor.shape=cross,decor.filled=shaded,decor.size=3thick}{v5}
\end{fmfgraph*}}}
\hspace*{3mm}
\parbox{10mm}{\centerline{
\begin{fmfgraph*}(10,10)
\setval
\fmfforce{1/2w,1h}{v1}
\fmfforce{0w,1/2h}{v2}
\fmfforce{1/2w,1/2h}{v3}
\fmfforce{1w,1/2h}{v4}
\fmfforce{1/2w,0h}{v5}
\fmf{plain}{v1,v5}
\fmf{plain}{v2,v4}
\fmfdot{v3}
\fmfv{decor.shape=cross,decor.filled=shaded,decor.size=3thick}{v1}
\fmfv{decor.shape=cross,decor.filled=shaded,decor.size=3thick}{v2}
\fmfv{decor.shape=cross,decor.filled=shaded,decor.size=3thick}{v4}
\fmfv{decor.shape=cross,decor.filled=shaded,decor.size=3thick}{v5}
\end{fmfgraph*}}}
\hspace*{3mm} + 16 \hspace*{3mm}
\parbox{15mm}{\centerline{
\begin{fmfgraph*}(15,10)
\setval
\fmfforce{1/3w,1h}{v1}
\fmfforce{2/3w,1h}{v2}
\fmfforce{0w,1/2h}{v3}
\fmfforce{1/3w,1/2h}{v4}
\fmfforce{2/3w,1/2h}{v5}
\fmfforce{1w,1/2h}{v6}
\fmfforce{1/3w,0h}{v7}
\fmfforce{2/3w,0h}{v8}
\fmf{plain}{v3,v6}
\fmf{plain}{v1,v7}
\fmf{plain}{v2,v8}
\fmfdot{v4}
\fmfdot{v5}
\fmfv{decor.shape=cross,decor.filled=shaded,decor.size=3thick}{v1}
\fmfv{decor.shape=cross,decor.filled=shaded,decor.size=3thick}{v2}
\fmfv{decor.shape=cross,decor.filled=shaded,decor.size=3thick}{v3}
\fmfv{decor.shape=cross,decor.filled=shaded,decor.size=3thick}{v6}
\fmfv{decor.shape=cross,decor.filled=shaded,decor.size=3thick}{v7}
\fmfv{decor.shape=cross,decor.filled=shaded,decor.size=3thick}{v8}
\end{fmfgraph*}}}
\hspace*{0.3cm} + {} 12 \hspace*{3mm}
\parbox{10mm}{\centerline{
\begin{fmfgraph*}(10,10)
\setval
\fmfforce{1/2w,1h}{v1}
\fmfforce{0w,1/2h}{v2}
\fmfforce{1/2w,1/2h}{v3}
\fmfforce{1w,1/2h}{v4}
\fmf{plain}{v2,v4}
\fmf{plain,left=1}{v3,v1,v3}
\fmfdot{v3}
\fmfv{decor.shape=cross,decor.filled=shaded,decor.size=3thick}{v2}
\fmfv{decor.shape=cross,decor.filled=shaded,decor.size=3thick}{v4}
\end{fmfgraph*}}}
\hspace*{3mm}
\parbox{10mm}{\centerline{
\begin{fmfgraph*}(10,10)
\setval
\fmfforce{1/2w,1h}{v1}
\fmfforce{0w,1/2h}{v2}
\fmfforce{1/2w,1/2h}{v3}
\fmfforce{1w,1/2h}{v4}
\fmfforce{1/2w,0h}{v5}
\fmf{plain}{v1,v5}
\fmf{plain}{v2,v4}
\fmfdot{v3}
\fmfv{decor.shape=cross,decor.filled=shaded,decor.size=3thick}{v1}
\fmfv{decor.shape=cross,decor.filled=shaded,decor.size=3thick}{v2}
\fmfv{decor.shape=cross,decor.filled=shaded,decor.size=3thick}{v4}
\fmfv{decor.shape=cross,decor.filled=shaded,decor.size=3thick}{v5}
\end{fmfgraph*}}}
\hspace*{3mm} + 72 \hspace*{1mm}
\parbox{15mm}{\centerline{
\begin{fmfgraph*}(15,10)
\setval
\fmfforce{0.0976w,0.85355h}{v1}
\fmfforce{0.902w,0.85355h}{v2}
\fmfforce{1/3w,1/2h}{v3}
\fmfforce{2/3w,1/2h}{v4}
\fmfforce{0.0976w,0.146h}{v5}
\fmfforce{0.902w,0.146h}{v6}
\fmf{plain}{v1,v3}
\fmf{plain}{v2,v4}
\fmf{plain}{v5,v3}
\fmf{plain}{v4,v6}
\fmf{plain,left=1}{v3,v4,v3}
\fmfdot{v3}
\fmfdot{v4}
\fmfv{decor.shape=cross,decor.filled=shaded,decor.size=3thick}{v1}
\fmfv{decor.shape=cross,decor.filled=shaded,decor.size=3thick}{v2}
\fmfv{decor.shape=cross,decor.filled=shaded,decor.size=3thick}{v5}
\fmfv{decor.shape=cross,decor.filled=shaded,decor.size=3thick}{v6}
\end{fmfgraph*}}}
\nonumber \\  &&
\hspace*{1cm} 
+ 36 \hspace{3mm}
\parbox{10mm}{\centerline{
\begin{fmfgraph*}(10,10)
\setval
\fmfforce{1/2w,1h}{v1}
\fmfforce{0w,1/2h}{v2}
\fmfforce{1/2w,1/2h}{v3}
\fmfforce{1w,1/2h}{v4}
\fmf{plain}{v2,v4}
\fmf{plain,left=1}{v3,v1,v3}
\fmfdot{v3}
\fmfv{decor.shape=cross,decor.filled=shaded,decor.size=3thick}{v2}
\fmfv{decor.shape=cross,decor.filled=shaded,decor.size=3thick}{v4}
\end{fmfgraph*}}}
\hspace*{3mm}
\parbox{10mm}{\centerline{
\begin{fmfgraph*}(10,10)
\setval
\fmfforce{1/2w,1h}{v1}
\fmfforce{0w,1/2h}{v2}
\fmfforce{1/2w,1/2h}{v3}
\fmfforce{1w,1/2h}{v4}
\fmf{plain}{v2,v4}
\fmf{plain,left=1}{v3,v1,v3}
\fmfdot{v3}
\fmfv{decor.shape=cross,decor.filled=shaded,decor.size=3thick}{v2}
\fmfv{decor.shape=cross,decor.filled=shaded,decor.size=3thick}{v4}
\end{fmfgraph*}}}
\hspace*{0.3cm} 
+ {} 96 \hspace*{3mm}
\parbox{15mm}{\centerline{
\begin{fmfgraph*}(15,10)
\setval
\fmfforce{1/3w,1h}{v1}
\fmfforce{2/3w,1h}{v2}
\fmfforce{0w,1/2h}{v3}
\fmfforce{1/3w,1/2h}{v4}
\fmfforce{2/3w,1/2h}{v5}
\fmfforce{1w,1/2h}{v6}
\fmfforce{1/3w,0h}{v7}
\fmf{plain}{v1,v7}
\fmf{plain}{v3,v6}
\fmf{plain,left=1}{v5,v2,v5}
\fmfdot{v4}
\fmfdot{v5}
\fmfv{decor.shape=cross,decor.filled=shaded,decor.size=3thick}{v1}
\fmfv{decor.shape=cross,decor.filled=shaded,decor.size=3thick}{v3}
\fmfv{decor.shape=cross,decor.filled=shaded,decor.size=3thick}{v6}
\fmfv{decor.shape=cross,decor.filled=shaded,decor.size=3thick}{v7}
\end{fmfgraph*}}}
\hspace*{0.3cm} + 6 \hspace*{3mm}
\parbox{10mm}{\centerline{
\begin{fmfgraph*}(10,10)
\setval
\fmfforce{0w,1/2h}{v1}
\fmfforce{1/2w,1/2h}{v2}
\fmfforce{1w,1/2h}{v3}
\fmf{plain,left=1}{v1,v2,v1}
\fmf{plain,left=1}{v2,v3,v2}
\fmfdot{v2}
\end{fmfgraph*}}}
\hspace*{3mm}
\parbox{10mm}{\centerline{
\begin{fmfgraph*}(10,10)
\setval
\fmfforce{1/2w,1h}{v1}
\fmfforce{0w,1/2h}{v2}
\fmfforce{1/2w,1/2h}{v3}
\fmfforce{1w,1/2h}{v4}
\fmfforce{1/2w,0h}{v5}
\fmf{plain}{v1,v5}
\fmf{plain}{v2,v4}
\fmfdot{v3}
\fmfv{decor.shape=cross,decor.filled=shaded,decor.size=3thick}{v1}
\fmfv{decor.shape=cross,decor.filled=shaded,decor.size=3thick}{v2}
\fmfv{decor.shape=cross,decor.filled=shaded,decor.size=3thick}{v4}
\fmfv{decor.shape=cross,decor.filled=shaded,decor.size=3thick}{v5}
\end{fmfgraph*}}}
\hspace*{3mm} +96 \hspace*{3mm}
\parbox{15mm}{\centerline{
\begin{fmfgraph*}(15,10)
\setval
\fmfforce{0w,1/2h}{v1}
\fmfforce{1/3w,1/2h}{v2}
\fmfforce{2/3w,1/2h}{v3}
\fmfforce{1w,1/2h}{v4}
\fmf{plain}{v1,v4}
\fmf{plain,left=1}{v2,v3,v2}
\fmfdot{v2}
\fmfdot{v3}
\fmfv{decor.shape=cross,decor.filled=shaded,decor.size=3thick}{v1}
\fmfv{decor.shape=cross,decor.filled=shaded,decor.size=3thick}{v4}
\end{fmfgraph*}}}
\hspace*{0.3cm} + {} 144 \hspace*{3mm}
\parbox{20mm}{\centerline{
\begin{fmfgraph*}(20,10)
\setval
\fmfforce{1/4w,1h}{v1}
\fmfforce{3/4w,1h}{v2}
\fmfforce{0w,1/2h}{v3}
\fmfforce{1/4w,1/2h}{v4}
\fmfforce{3/4w,1/2h}{v5}
\fmfforce{1w,1/2h}{v6}
\fmf{plain}{v3,v6}
\fmf{plain,left=1}{v4,v1,v4}
\fmf{plain,left=1}{v5,v2,v5}
\fmfdot{v4}
\fmfdot{v5}
\fmfv{decor.shape=cross,decor.filled=shaded,decor.size=3thick}{v3}
\fmfv{decor.shape=cross,decor.filled=shaded,decor.size=3thick}{v6}
\end{fmfgraph*}}}
\nonumber \\ [2mm] &&
\hspace*{1cm} 
+ 36 \hspace*{3mm}
\parbox{10mm}{\centerline{
\begin{fmfgraph*}(10,10)
\setval
\fmfforce{0w,1/2h}{v1}
\fmfforce{1/2w,1/2h}{v2}
\fmfforce{1w,1/2h}{v3}
\fmf{plain,left=1}{v1,v2,v1}
\fmf{plain,left=1}{v2,v3,v2}
\fmfdot{v2}
\end{fmfgraph*}}}
\hspace*{3mm}
\parbox{10mm}{\centerline{
\begin{fmfgraph*}(10,10)
\setval
\fmfforce{1/2w,1h}{v1}
\fmfforce{0w,1/2h}{v2}
\fmfforce{1/2w,1/2h}{v3}
\fmfforce{1w,1/2h}{v4}
\fmf{plain}{v2,v4}
\fmf{plain,left=1}{v3,v1,v3}
\fmfdot{v3}
\fmfv{decor.shape=cross,decor.filled=shaded,decor.size=3thick}{v2}
\fmfv{decor.shape=cross,decor.filled=shaded,decor.size=3thick}{v4}
\end{fmfgraph*}}}
\hspace*{3mm} + 144 \hspace*{1mm}
\parbox{15mm}{\centerline{
\begin{fmfgraph*}(15,10)
\setval
\fmfforce{0.0976w,0.85355h}{v1}
\fmfforce{1/3w,1/2h}{v2}
\fmfforce{2/3w,1/2h}{v3}
\fmfforce{1w,1/2h}{v4}
\fmfforce{0.0976w,0.146h}{v5}
\fmf{plain}{v1,v2}
\fmf{plain}{v5,v2}
\fmf{plain,left=1}{v2,v3,v2}
\fmf{plain,left=1}{v3,v4,v3}
\fmfdot{v2}
\fmfdot{v3}
\fmfv{decor.shape=cross,decor.filled=shaded,decor.size=3thick}{v1}
\fmfv{decor.shape=cross,decor.filled=shaded,decor.size=3thick}{v5}
\end{fmfgraph*}}}
\hspace*{0.3cm} + {} 72 \hspace*{3mm}
\parbox{15mm}{\centerline{
\begin{fmfgraph*}(15,10)
\setval
\fmfforce{0w,1/2h}{v1}
\fmfforce{1/3w,1/2h}{v2}
\fmfforce{2/3w,1/2h}{v3}
\fmfforce{1w,1/2h}{v4}
\fmf{plain,left=1}{v1,v2,v1}
\fmf{plain,left=1}{v2,v3,v2}
\fmf{plain,left=1}{v3,v4,v3}
\fmfdot{v2}
\fmfdot{v3}
\end{fmfgraph*}}}
\hspace*{3mm}+ 24 \hspace*{3mm}
\parbox{10mm}{\centerline{
\begin{fmfgraph*}(10,10)
\setval
\fmfforce{0w,1/2h}{v1}
\fmfforce{1w,1/2h}{v2}
\fmf{plain,left=1}{v1,v2,v1}
\fmf{plain,left=1/2}{v1,v2}
\fmf{plain,left=1/2}{v2,v1}
\fmfdot{v1}
\fmfdot{v2}
\end{fmfgraph*}}}
\hspace*{3mm} + 9 \hspace*{3mm}
\parbox{10mm}{\centerline{
\begin{fmfgraph*}(10,10)
\setval
\fmfforce{0w,1/2h}{v1}
\fmfforce{1/2w,1/2h}{v2}
\fmfforce{1w,1/2h}{v3}
\fmf{plain,left=1}{v1,v2,v1}
\fmf{plain,left=1}{v2,v3,v2}
\fmfdot{v2}
\end{fmfgraph*}}}
\hspace*{3mm}
\parbox{10mm}{\centerline{
\begin{fmfgraph*}(10,10)
\setval
\fmfforce{0w,1/2h}{v1}
\fmfforce{1/2w,1/2h}{v2}
\fmfforce{1w,1/2h}{v3}
\fmf{plain,left=1}{v1,v2,v1}
\fmf{plain,left=1}{v2,v3,v2}
\fmfdot{v2}
\end{fmfgraph*}}}
\hspace*{3mm} . \label{HJK}
%
\end{eqnarray}
We observe that
contributions from both connected
and disconnected Feynman diagrams appear.
The disconnected diagrams
vanish once we rewrite the
imaginary-time evolution amplitude\index{time evolution amplitude}
in the form
\begin{eqnarray}
\label{ehochw}
(x_b \,  \hbar \beta | x_a \, 0) = 
(x_b \, \hbar \beta | x_a  \, 0 )_{\omega} 
\exp \left[ W (x_b, \hbar \beta; x_a,0) \right]
\, ,
\end{eqnarray}
where the
exponent $W (x_b, \hbar \beta; x_a,0)$ contains only the connected
Feynman diagrams. We obtain from (\ref{2B}) and 
(\ref{firstorderfeynman})--(\ref{ehochw}) 
the expansion
\begin{eqnarray}
\label{secondorderdensitymatrixfeynman}
W ( x_b, \hbar \beta ; x_a , 0) & = &
- \frac{g}{\hbar} \left( \hspace*{2mm}
%
%
%
%
%
%
\parbox{10mm}{\centerline{
\begin{fmfgraph*}(10,10)
\setval
\fmfforce{1/2w,1h}{v1}
\fmfforce{0w,1/2h}{v2}
\fmfforce{1/2w,1/2h}{v3}
\fmfforce{1w,1/2h}{v4}
\fmfforce{1/2w,0h}{v5}
\fmf{plain}{v1,v5}
\fmf{plain}{v2,v4}
\fmfdot{v3}
\fmfv{decor.shape=cross,decor.filled=shaded,decor.size=3thick}{v1}
\fmfv{decor.shape=cross,decor.filled=shaded,decor.size=3thick}{v2}
\fmfv{decor.shape=cross,decor.filled=shaded,decor.size=3thick}{v4}
\fmfv{decor.shape=cross,decor.filled=shaded,decor.size=3thick}{v5}
\end{fmfgraph*}}}
\hspace*{3mm} + 6 \hspace*{3mm}
\parbox{10mm}{\centerline{
\begin{fmfgraph*}(10,10)
\setval
\fmfforce{1/2w,1h}{v1}
\fmfforce{0w,1/2h}{v2}
\fmfforce{1/2w,1/2h}{v3}
\fmfforce{1w,1/2h}{v4}
\fmf{plain}{v2,v4}
\fmf{plain,left=1}{v3,v1,v3}
\fmfdot{v3}
\fmfv{decor.shape=cross,decor.filled=shaded,decor.size=3thick}{v2}
\fmfv{decor.shape=cross,decor.filled=shaded,decor.size=3thick}{v4}
\end{fmfgraph*}}}
\hspace*{3mm} + 3 \hspace*{3mm}
\parbox{10mm}{\centerline{
\begin{fmfgraph*}(10,10)
\setval
\fmfforce{0w,1/2h}{v1}
\fmfforce{1/2w,1/2h}{v2}
\fmfforce{1w,1/2h}{v3}
\fmf{plain,left=1}{v1,v2,v1}
\fmf{plain,left=1}{v2,v3,v2}
\fmfdot{v2}
\end{fmfgraph*}}}
\hspace*{3mm}
\right) 
\hspace*{0.2cm} + {} \frac{g^2}{2 \hbar^2} \left(
%
%
%
%
%
%
%
%
%
8 \hspace*{3mm}
\parbox{15mm}{\centerline{
\begin{fmfgraph*}(15,10)
\setval
\fmfforce{1/3w,1h}{v1}
\fmfforce{2/3w,1h}{v2}
\fmfforce{0w,1/2h}{v3}
\fmfforce{1/3w,1/2h}{v4}
\fmfforce{2/3w,1/2h}{v5}
\fmfforce{1w,1/2h}{v6}
\fmfforce{1/3w,0h}{v7}
\fmfforce{2/3w,0h}{v8}
\fmf{plain}{v3,v6}
\fmf{plain}{v1,v7}
\fmf{plain}{v2,v8}
\fmfdot{v4}
\fmfdot{v5}
\fmfv{decor.shape=cross,decor.filled=shaded,decor.size=3thick}{v1}
\fmfv{decor.shape=cross,decor.filled=shaded,decor.size=3thick}{v2}
\fmfv{decor.shape=cross,decor.filled=shaded,decor.size=3thick}{v3}
\fmfv{decor.shape=cross,decor.filled=shaded,decor.size=3thick}{v6}
\fmfv{decor.shape=cross,decor.filled=shaded,decor.size=3thick}{v7}
\fmfv{decor.shape=cross,decor.filled=shaded,decor.size=3thick}{v8}
\end{fmfgraph*}}}
\hspace*{3mm} + 36 \hspace*{1mm}
\parbox{15mm}{\centerline{
\begin{fmfgraph*}(15,10)
\setval
\fmfforce{0.0976w,0.85355h}{v1}
\fmfforce{0.902w,0.85355h}{v2}
\fmfforce{1/3w,1/2h}{v3}
\fmfforce{2/3w,1/2h}{v4}
\fmfforce{0.0976w,0.146h}{v5}
\fmfforce{0.902w,0.146h}{v6}
\fmf{plain}{v1,v3}
\fmf{plain}{v2,v4}
\fmf{plain}{v5,v3}
\fmf{plain}{v4,v6}
\fmf{plain,left=1}{v3,v4,v3}
\fmfdot{v3}
\fmfdot{v4}
\fmfv{decor.shape=cross,decor.filled=shaded,decor.size=3thick}{v1}
\fmfv{decor.shape=cross,decor.filled=shaded,decor.size=3thick}{v2}
\fmfv{decor.shape=cross,decor.filled=shaded,decor.size=3thick}{v5}
\fmfv{decor.shape=cross,decor.filled=shaded,decor.size=3thick}{v6}
\end{fmfgraph*}}}
\right. 
\nonumber \\ [2mm]
& & \left.
+ 48 \hspace{3mm}
\parbox{15mm}{\centerline{
\begin{fmfgraph*}(15,10)
\setval
\fmfforce{1/3w,1h}{v1}
\fmfforce{2/3w,1h}{v2}
\fmfforce{0w,1/2h}{v3}
\fmfforce{1/3w,1/2h}{v4}
\fmfforce{2/3w,1/2h}{v5}
\fmfforce{1w,1/2h}{v6}
\fmfforce{1/3w,0h}{v7}
\fmf{plain}{v1,v7}
\fmf{plain}{v3,v6}
\fmf{plain,left=1}{v5,v2,v5}
\fmfdot{v4}
\fmfdot{v5}
\fmfv{decor.shape=cross,decor.filled=shaded,decor.size=3thick}{v1}
\fmfv{decor.shape=cross,decor.filled=shaded,decor.size=3thick}{v3}
\fmfv{decor.shape=cross,decor.filled=shaded,decor.size=3thick}{v6}
\fmfv{decor.shape=cross,decor.filled=shaded,decor.size=3thick}{v7}
\end{fmfgraph*}}}
\right.
{}+ 48 \hspace*{3mm}
\parbox{15mm}{\centerline{
\begin{fmfgraph*}(15,10)
\setval
\fmfforce{0w,1/2h}{v1}
\fmfforce{1/3w,1/2h}{v2}
\fmfforce{2/3w,1/2h}{v3}
\fmfforce{1w,1/2h}{v4}
\fmf{plain}{v1,v4}
\fmf{plain,left=1}{v2,v3,v2}
\fmfdot{v2}
\fmfdot{v3}
\fmfv{decor.shape=cross,decor.filled=shaded,decor.size=3thick}{v1}
\fmfv{decor.shape=cross,decor.filled=shaded,decor.size=3thick}{v4}
\end{fmfgraph*}}}
\hspace*{3mm} + 72 \hspace*{3mm}
\parbox{20mm}{\centerline{
\begin{fmfgraph*}(20,10)
\setval
\fmfforce{1/4w,1h}{v1}
\fmfforce{3/4w,1h}{v2}
\fmfforce{0w,1/2h}{v3}
\fmfforce{1/4w,1/2h}{v4}
\fmfforce{3/4w,1/2h}{v5}
\fmfforce{1w,1/2h}{v6}
\fmf{plain}{v3,v6}
\fmf{plain,left=1}{v4,v1,v4}
\fmf{plain,left=1}{v5,v2,v5}
\fmfdot{v4}
\fmfdot{v5}
\fmfv{decor.shape=cross,decor.filled=shaded,decor.size=3thick}{v3}
\fmfv{decor.shape=cross,decor.filled=shaded,decor.size=3thick}{v6}
\end{fmfgraph*}}}
\hspace*{3mm} + 72 \hspace*{1mm}
\parbox{15mm}{\centerline{
\begin{fmfgraph*}(15,10)
\setval
\fmfforce{0.0976w,0.85355h}{v1}
\fmfforce{1/3w,1/2h}{v2}
\fmfforce{2/3w,1/2h}{v3}
\fmfforce{1w,1/2h}{v4}
\fmfforce{0.0976w,0.146h}{v5}
\fmf{plain}{v1,v2}
\fmf{plain}{v5,v2}
\fmf{plain,left=1}{v2,v3,v2}
\fmf{plain,left=1}{v3,v4,v3}
\fmfdot{v2}
\fmfdot{v3}
\fmfv{decor.shape=cross,decor.filled=shaded,decor.size=3thick}{v1}
\fmfv{decor.shape=cross,decor.filled=shaded,decor.size=3thick}{v5}
\end{fmfgraph*}}}
\nonumber \\ [2mm]
& & 
\left.
{}+ 36 \hspace*{3mm}
\parbox{15mm}{\centerline{
\begin{fmfgraph*}(15,10)
\setval
\fmfforce{0w,1/2h}{v1}
\fmfforce{1/3w,1/2h}{v2}
\fmfforce{2/3w,1/2h}{v3}
\fmfforce{1w,1/2h}{v4}
\fmf{plain,left=1}{v1,v2,v1}
\fmf{plain,left=1}{v2,v3,v2}
\fmf{plain,left=1}{v3,v4,v3}
\fmfdot{v2}
\fmfdot{v3}
\end{fmfgraph*}}}
\hspace*{3mm} + 12 \hspace*{3mm}
\parbox{10mm}{\centerline{
\begin{fmfgraph*}(10,10)
\setval
\fmfforce{0w,1/2h}{v1}
\fmfforce{1w,1/2h}{v2}
\fmf{plain,left=1}{v1,v2,v1}
\fmf{plain,left=1/2}{v1,v2}
\fmf{plain,left=1/2}{v2,v1}
\fmfdot{v1}
\fmfdot{v2}
\end{fmfgraph*}}}
\hspace*{3mm}
\right) + ...  \, ,
\end{eqnarray}
where disconnected diagrams are indeed no longer present.
As mentioned above, we restrict ourselves to the low-temperature limit
of the diagonal elements of the density matrix
which determine the ground-state wave function.
In order to evaluate the various contributions in
(\ref{secondorderdensitymatrixfeynman}),
we need the classical path (\ref{xcl})
and the Green function\index{Green function}
(\ref{Green})
in the low-temperature limit:
\begin{eqnarray}
\label{XcllowT}
\lim_{\beta \rightarrow \infty } \, x_{\rm cl} (\tau) &=&
x \left( e^{- \omega \tau}
+ e^{-\omega (\hbar \beta - \tau)} \right) \, ,\\
\label{GlowT}
\lim_{\beta \rightarrow \infty } \, G (\tau_1, \tau_2) & = & 
\frac{\hbar}{2 M \omega} \left[ \,\,
\theta (\tau_1 - \tau_2) \, e^{- \omega (\tau_1 - \tau_2)} 
+ \theta (\tau_2 - \tau_1) \, e^{- \omega (\tau_2 - \tau_1) } 
- e^{- \omega (\tau_1 + \tau_2)}
- e^{-2 \hbar \beta \omega + \omega (\tau_1 + \tau_2)} 
\phantom{e^1} \hspace*{-0.2cm}\right] \, .
\end{eqnarray}
Computing with these expressions
the Feynman diagrams\index{Feynman diagrams}
in (\ref{secondorderdensitymatrixfeynman}),
the low-temperature limit of the imaginary-time evolution amplitude
(\ref{ehochw}) reads together with (\ref{harmonic2}) 
\begin{eqnarray}
\label{zwischenergebnis}
\lim_{\beta \rightarrow \infty} 
(x \, \hbar \beta | x \, 0) &=&
\lim_{\beta \rightarrow \infty}
\sqrt{ \frac{M \omega}{ \hbar \pi } }
\exp \left[ - \frac{\hbar \beta \omega}{2}
- \frac{M \omega}{\hbar} x^2 
+  \frac{g}{\hbar} \left( \frac{9 \hbar^2}{8 M^2 \omega^3}
- \frac{3 \hbar^3 \beta}{4 M^2 \omega^2}
- \frac{3 \hbar}{2 M \omega^2} x^2
- \frac{1}{2 \omega} x^4 \right) 
\right. \nonumber \\ &&  \left. 
+ \frac{g^2}{2 \hbar^2} \left( \frac{21 \hbar^5 \beta}{4 M^4 \omega^5} 
- \frac{205 \hbar^4}{16 M^4 \omega^6} 
+ \frac{21 \hbar^3}{2 M^3 \omega^5} x^2
+ \frac{11 \hbar^2}{4 M^2 \omega^4} x^4
+ \frac{\hbar}{3 M \omega^3} x^6
\right) + ... 
\right]  \, .
\end{eqnarray}
According to (\ref{partitionfunction}),
the partition function $Z$ follows from (\ref{zwischenergebnis}) by
performing an integration with respect to $x$. 
This results in
\begin{eqnarray}
\label{AHOpartitionfunction}
\lim_{\beta \rightarrow \infty} Z = \lim_{\beta \rightarrow \infty}
\exp \left( - \frac{\hbar \beta \omega}{2}
- \frac{3 g \hbar^2 \beta}{4 M^2 \omega^2}
+ \frac{21 g^2 \hbar^3 \beta}{8 M^4 \omega^5} 
 + ... \right) \, .
\end{eqnarray}
Inserting (\ref{zwischenergebnis}) and
(\ref{AHOpartitionfunction}) into (\ref{densitymatrix})
we observe a cancellation of all 
terms which would diverge in the
low-temperature limit $\beta \rightarrow \infty$.
Thus the diagonal elements of the density matrix read in this limit
\begin{eqnarray}
\label{rhoexp}
\lim_{\beta \rightarrow \infty}
\rho (x,x) &=& \sqrt{ \frac{M \omega}{\hbar \pi} }
\exp \left[ - \frac{M \omega}{\hbar} x^2 
+ \frac{g}{\hbar} \left( 
\frac{9 \hbar^2}{8 M^2 \omega^3}
- \frac{3 \hbar}{2 M \omega^2} x^2
- \frac{1}{ 2 \omega} x^4
\right) \right. 
\nonumber \\ & &  \left.
+ \frac{g^2}{2 \hbar^2} \left(
- \frac{205 \hbar^4}{16 M^4 \omega^6}
+ \frac{21 \hbar^4}{2 M^3 \omega^5} x^2 
+ \frac{11 \hbar^2}{4 M^2 \omega^4} x^4
+ \frac{\hbar}{3 M \omega^3} x^6
\right) + ...
\right] \, .
\end{eqnarray}
By taking the square root and expanding the exponential
term up to second order in the coupling strength $g$,
we derive the second-order ground-state wave
function\index{wave function} (\ref{wavefunction}):
\begin{eqnarray}
\label{expandedwave}
\Psi (x) &= &\left( \frac{M \omega}{\hbar \pi} \right)^{1/4} \,
\exp \left( - \frac{M \omega}{2 \hbar} x^2 \right) 
\left[
1 - \frac{g}{\hbar}
\left(
- \frac{9 \hbar^2}{16 M^2 \omega^3}
+ \frac{3 \hbar}{4 M \omega^2} x^2
+ \frac{1}{4 \omega} x^4 \right)  
\right. \nonumber \\ & & \left. 
+ \frac{g^2}{2 \hbar^2}\left(
-  \frac{1559 \hbar^4}{256 M^4 \omega^6}
+ \frac{141 \hbar^3}{32 M^3 \omega^5} x^2
+ \frac{53 \hbar^2}{32 M^2 \omega^4} x^4
+ \frac{13 \hbar}{24 M \omega^3} x^6
+ \frac{1}{16 \omega^2} x^8
\right) + ...
\right] \, .
\end{eqnarray}
This result corresponds to the solution of the
Bender-Wu\index{BENDER, X.} \index{WU, X.}
recursion relation \cite{Bender/Wu} for the ground-state
wave function which is normalized such that
\begin{eqnarray}
\label{normalization1}
\int_{- \infty}^{+ \infty} dx \, \Psi^2 (x) = 1
\end{eqnarray}
holds up to second order in the coupling strength $g$.
\section{Variational~Perturbation~Theory\index{variational perturbation 
theory}}
Variational perturbation theory enables us
to evaluate the ground-state
wave function\index{wave function} for all values of the coupling
constant $g$ and
even in the strong-coupling limit\index{strong-coupling limit}
$g \rightarrow \infty$.
To this end
we simply add and subtract
a harmonic oscillator\index{harmonic oscillator} of
trial frequency\index{trial frequency} $\Omega$ to the anharmonic oscillator
potential (\ref{AHOpotential}):
\begin{eqnarray}
\label{trick}
V (x) & = &
\frac{M}{2} \Omega^2 x^2 + g \frac{M}{2} \frac{\omega^2 - \Omega^2}{g} \, x^2 + g \, x^4 \, .
\end{eqnarray}
We now treat the second term as if it was of the order of 
the coupling constant $g$.
The result is obtained most simply by
substituting the frequency $\omega$
in the original anharmonic oscillator potential
(\ref{AHOpotential}) according to Kleinert's trick \cite{Kleinert}
\begin{eqnarray}
\label{wurzeltrick}
\omega \rightarrow \Omega \sqrt{ 1 + gr } \, ,
\end{eqnarray}
where we have
\begin{eqnarray}
\label{gr}
r \equiv \frac{\omega^2 - \Omega^2}{g \Omega^2} \, .
\end{eqnarray}
Writing the ground-state wave function (\ref{expandedwave})
in the form $\Psi(x)=\exp [W(x)]$
with the cumulant expansion\index{cumulant expansion}
\begin{eqnarray}
W(x)  &=&
\exp \left[
\frac{1}{4} \log \left( \frac{M \omega}{\hbar \pi} \right)
- \frac{M \omega}{2 \hbar} x^2
+ \frac{g}{\hbar} \left( 
\frac{9 \hbar^2}{16 M^2 \omega^3}
- \frac{3 \hbar}{4 M \omega^2} x^2
- \frac{1}{ 16 \omega} x^4
\right) \right. \nonumber \\
& & \left.
+ \frac{g^2}{2 \hbar^2} \left(
- \frac{205 \hbar^4}{32 M^4 \omega^6}
+ \frac{21 \hbar^4}{4 M^3 \omega^5} x^2
+ \frac{11 \hbar^2}{8 M^2 \omega^4} x^4
+ \frac{\hbar}{6 M \omega^3} x^6
\right) + ...
\right] \, ,
\label{c}
\end{eqnarray}
we apply the trick (\ref{wurzeltrick})
to $W(x)$, and reexpand in powers of $g$ at fixed $r$.
Afterwards $r$ is substituted according to (\ref{gr}).
Thus we obtain in the first order
\begin{eqnarray}
\label{waveinexp1}
\Psi^{(1)} (x, \Omega) & = & \exp \left[ \frac{1}{4} \log \left( 
\frac{M \Omega}{\hbar \pi} \right)
- \frac{1}{8} + \frac{\omega^2}{8 \Omega^2}
 - \frac{M \Omega}{4 \hbar} \left( 1+ \frac{\omega^2}{\Omega^2} \right) x^2
\right. \nonumber \\
& & \left.
+ \frac{g}{\hbar} \left( \frac{9 \hbar^2}{16 M^2 \Omega^3}
- \frac{3 \hbar}{4 M \Omega^2} x^2
- \frac{1}{4 \Omega} x^4 \right) \right] \, ,
\end{eqnarray}
whereas the second-order expansion reads
\begin{eqnarray}
\Psi^{(2)}(x,\Omega) &= &
\exp \left\{ \frac{1}{4} \log \left( \frac{M \Omega}{\hbar \pi} \right)
- \frac{1}{16} + \frac{\omega^2}{4 \Omega^2}
- \frac{\omega^4}{16 \Omega^4} 
- \frac{M \Omega}{2 \hbar} \left( \frac{3}{8} + \frac{3 \omega^2}{4 \Omega^2}
- \frac{\omega^4}{8 \Omega^4} \right) x^2
\right. \nonumber \\& & \left.
+ \frac{g}{\hbar} \left[
\frac{9 \hbar^2}{ 16 M^2 \Omega^3} \left(  \frac{5}{2} - \frac{3 
\omega^2}{2 \Omega^2}  \right) 
- \frac{3 \hbar}{4 M \Omega^2} \left( 2 - \frac{\omega^2}{\Omega^2}  
\right) x^2
- \frac{1}{4 \Omega} \left(  \frac{3}{2} -  \frac{\omega^2}{2 \Omega^2} 
\right) x^4 \right] \right.
\nonumber \\& & \left.
+ \frac{g^2}{2 \hbar^2} \left[
- \frac{205 \hbar^4}{32 M^4 \Omega^6}
+ \frac{21 \hbar^3}{2 M^3 \Omega^5} x^2 
+ \frac{11 \hbar^2}{8 M^2 \Omega^4} x^4
+ \frac{\hbar}{6 M \Omega^3} x^6
\right] \right\} \, .
\label{waveinexp}
\end{eqnarray}
Both in first and in second order,
the ground-state wave function\index{wave function}
depends on the artificially introduced frequency
parameter $\Omega$. According to the principle of
minimal sensitivity\index{principle of minimal sensitivity}
\cite{Stevenson}
we minimize its influence on $\Psi^{(n)} (x,\Omega)$ by searching for local
extrema of
$\Psi^{(n)}(x,\Omega)$ with respect to $\Omega$. As we have written the
wave function\index{wave function} in the form
$\Psi^{(n)}(x,\Omega)=\exp [W^{(n)}(x,\Omega)]$, it is sufficient
to take into account just
the inner derivative of $\Psi^{(n)}(x, \Omega)$,
i.e. we obtain the condition
$\partial W^{(n)}(x,\Omega)/\partial \Omega = 0$.

\begin{figure}[t]
\centerline{\epsfxsize=8cm \epsfbox{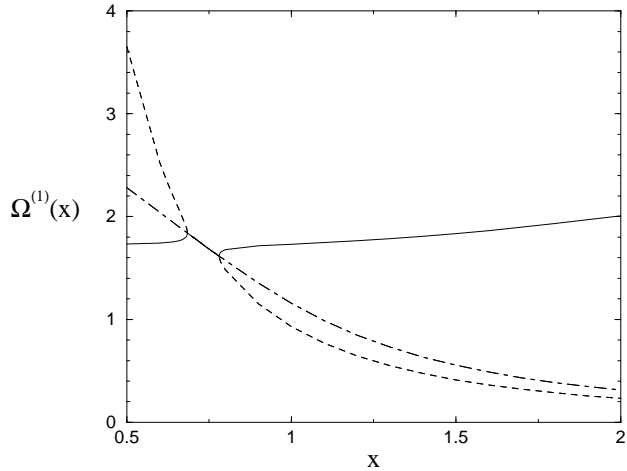}}
\caption{\label{fig2} First-order
results for the variational parameter
$\Omega$ at the intermediate coupling $g=1/2$.
The extremal branches 
for $x<0.684$ and for $x>0.780$ (solid lines and dashed lines)
are obtained from
the equation $\partial W^{(1)}(x,\Omega)/\partial \Omega =0$. 
For $0.684<x<0.780$ there are no real positive
solutions of this equation. 
Thus we look in this interval also
for turning points,
i.e. we determine real positive solutions of the
equation $\partial^2 W^{(1)}(x,\Omega) / \partial \Omega^2 =0$.
The curve for the turning points on the entire interval lies
between the two other branches (dot-dashed line) and
fills the gap.
Thus we can take those branches
into account which provide us with the most
continuous function $\Omega^{(1)} (x)$, i.e. the solid line.}
\end{figure}

It turns out in the first order $n=1$ that
this equation has two solutions for
$x<0.684$ and for $x>0.780$, however in the interval $0.684 < x < 0.780$,
$\Psi^{(1)} (x, \Omega)$ does not have any extremum \cite{Kunihiro}.
In accordance with the 
principle of minimal sensitivity\index{principle of minimal sensitivity}
we look for turning points on that
interval instead, i.e. we solve 
$\partial^2 W^{(1)}(x,\Omega) / \partial \Omega^2 =0$.
Fig.~\ref{fig2} shows how
the curve for the turning points links the extremal branches.
Now we have to choose which one of the branches of
$\Omega^{1}(x)$ we take into account. Inserting the lower branch for $x>0.780$
into the wave function (\ref{waveinexp1}) leads to unphysical results as
the ground-state wave function explodes dramatically.
Thus we choose the upper branch for $x>0.780$.
For $x<0.684$ the wave function becomes rather
independent of the choice of
$\Omega$. As we are looking for a function $\Omega^{(1)} (x)$ which is as
smooth as possible,
we choose the lower branch for $x<0.684$. 
Fig.~\ref{fig2} shows all branches of
$\Omega$ and highlights our final choice by a solid line.
\begin{figure}[t]
\centerline{\epsfxsize=8cm \epsfbox{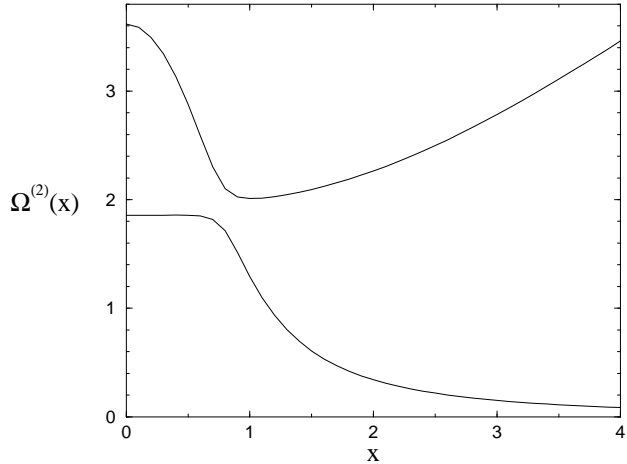}}
\caption{\label{fig1} The two positive branches of
$\Omega^{(2)} (x)$ on the interval $[0,4]$ for intermediate coupling $g=1/2$
obtained by solving the turning point equation
$\partial^2 W^{(2)} (x,\Omega) / \partial \Omega^2 = 0$.
In order to achieve the smoothest function we 
choose the lower branch for $x<0.8$ and
the upper branch for $x>0.8$. This
choice is justified by the results of our
first-order calculation (see Fig.~\ref{fig2}).}
\end{figure}

For the second order $n=2$
there are no real positive solutions of the equation
$\partial W^{(2)} (x,\Omega) / \partial \Omega = 0$
on the interval $x=[0,4]$.
Once more we have to look for turning
points instead and solve $\partial^2 W^{(2)}(x,\Omega)/\partial \Omega^2=0$.
This equation has two positive solutions
on this interval, so
we get two branches for the
solution $\Omega^{(2)}(x)$ (see Fig.~\ref{fig1}).
Again we have to choose one of these two branches.
Relying on a similar argument as for the first-order
we choose the upper branch for $x>0.8$ and the lower one for
$x<0.8$.
The perturbation series converges so quickly that the curves for the
first and second order as well as the exact
ground-state wave function are not
distinguishable on the plots. To see the difference
we determine the mean square deviation from the
exact numerical solution
\begin{eqnarray}
\label{meansqaredeviation}
D^{(n)} = 2 \int_0^{\infty} dx \left[
\Psi^{(n)} (x) - \Psi^{\rm ex} (x) \right]^2 \, ,
\end{eqnarray}
where the index $n$ denotes the order.
The integration is performed numerically. The factor 2 is introduced
for symmetry reasons, since we restrict our calculations to the positive
$x$-axis. It turns out that the mean square deviation
$D^{(2)} = 6.8 \times 10^{-7}$
is smaller than $D^{(1)} = 1.1 \times 10^{-5}$ by a
factor of $0.063$, which indicates
that variational perturbation theory converges very quickly also
for the ground-state wave function.
The same applies to both weak 
\begin{figure}[t]
\centerline{\epsfxsize=8cm \epsfbox{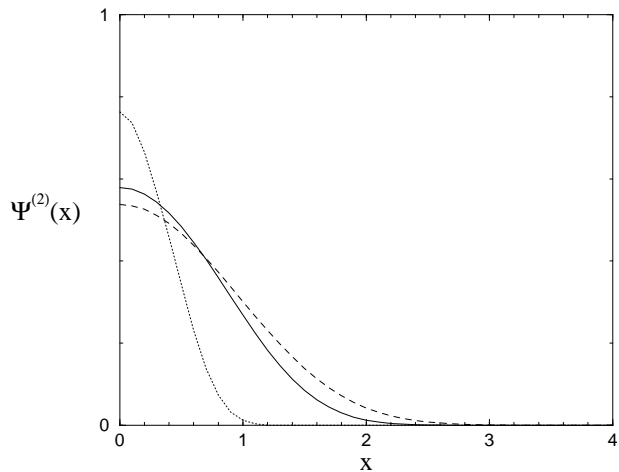}}
\caption{\label{fig4} The normalized second-order ground-state wave function
for weak coupling (dashed, $g=0.1$), for intermediate coupling
(solid, $g=1/2$), and for strong coupling (dotted, $g=50$).}
\end{figure}
and strong coupling as is illustrated in
Fig.~\ref{fig4} which shows the second-order ground-state
wave function $\Psi^{(2)} (x)$
for $g=0.1$, $g=1/2$, and $g=50$, respectively.

Note
that variational perturbation theory does not
preserve the normalization\index{normalization} of the
wave function. Although the perturbative ground-state
wave function (\ref{expandedwave}) is still
normalized in the usual sense,
this normalization is spoilt by extremizing
$\Psi (x, \Omega)$ with respect to the frequency parameter $\Omega$.
Thus we have to normalize the variational ground-state wave function
at the end.
%
%
\section*{Acknowledgements}
Both of us congratulate Professor Kleinert to his 60th Birthday.
F.W. thanks Professor Kleinert for supervising the research for
his diploma thesis where the results of this article were derived.
Moreover F.W. is very grateful to
his family for their patience and their support.
Finally both authors thank Michael Bachmann\index{BACHMANN, M.}
who always found time for fruitful discussions on
variational perturbation theory.

\end{fmffile}


\begin{thebibliography}{99}
%
\bibitem{Kleinert} 
H.~Kleinert\index{KLEINERT, H.}, 
{\it Path Integrals in Mechanics, Statistics, and Polymer Physics},
2nd ed.,
(World Scientific, Singapore, 1995).
%
\bibitem{Janke1}
W.~Janke\index{JANKE, W.} and
H.~Kleinert\index{KLEINERT, H.},
{\it Phys. Rev. Lett.} {\bf 75}, 2787 (1995).
%
\bibitem{Janke2}
H.~Kleinert\index{KLEINERT, H.} 
and W.~Janke\index{JANKE, W.},
{\it Phys. Lett. A} {\bf 206}, 283 (1995).
%
\bibitem{Kunihiro} 
T.~Kunihiro\index{KUNIHIRO, T.}, 
{\it Phys. Rev. Lett.} {\bf 78}, 3229 (1997).
%
\bibitem{Dichtematrix}
M.~Bachmann\index{BACHMANN, M.},
H.~Kleinert\index{KLEINERT. K.}, 
and A.~Pelster\index{PELSTER, A.},
{\it Phys. Rev. A} {\bf 60}, 3429 (1999).
%
\bibitem{KPB}
H.~Kleinert\index{KLEINERT, H.},
A.~Pelster\index{PELSTER. A}, and
M.~Bachmann\index{BACHMANN, M.},
{\it Phys. Rev. E} {\bf 60}, 2510 (1999).
%
\bibitem{Bender/Wu}
C.M.~Bender\index{BENDER, C.M.} and
T.T.~Wu\index{WU, T.T.},
{\it Phys. Rev.} {\bf 184}, 1231 (1969);
{\it Phys. Rev. D} {\bf 7}, 1620 (1973).
%
\bibitem{Stevenson}
P.M.~Stevenson\index{STEVENSON, P.M.},
{\it Phys. Rev. D} {\bf 23}, 2916 (1981).
%
\end{thebibliography}
\end{document}